\documentclass[aps,twocolumn,showpacs,floatfix,superscriptaddress]{revtex4-1}
\usepackage[linktocpage,bookmarksopen,bookmarksnumbered]{hyperref}
\usepackage[utf8]{inputenc} 
\usepackage{dcolumn}
\usepackage{ulem}
\usepackage{graphicx}
\usepackage{diagbox}
\usepackage{chemformula}
\usepackage{amsfonts}
\usepackage{amsmath,bm,amssymb}
\usepackage{blkarray}
\usepackage{braket} 
\usepackage{gensymb}

\begin{document}

\title{Calculating magnetic interactions in organic electrides}

\author{Taek Jung Kim}
\affiliation{Department of Physics, KAIST, 291 Daehak-ro, Yuseong-gu, Daejeon 34141, Republic of Korea }

\author{Hongkee Yoon} 
\affiliation{Department of Physics, KAIST, 291 Daehak-ro, Yuseong-gu, Daejeon 34141, Republic of Korea }

\author{Myung Joon Han} \email{mj.han@kaist.ac.kr}
\affiliation{Department of Physics, KAIST, 291 Daehak-ro, Yuseong-gu, Daejeon 34141, Republic of Korea }
\affiliation{ KAIST Institute for the NanoCentury, Korea Advanced
  Institute of Science and Technology, Daejeon 305-701, Korea }

\date{\today}

\begin{abstract}
We present our calculation results for organic magnetic electrides. In order to identify the ‘cavity’ electrons, we use maximally-localized Wannier functions and ‘empty atom’ technique. The estimation of magnetic coupling is then performed based on magnetic force linear response theory. Both short- and long-range magnetic interactions are calculated with a single self-consistent calculation of a primitive cell. With this scheme we investigate four different organic electrides whose magnetic properties have been partly unknown or under debate. Our calculation results unveil the nature of magnetic moment and their interactions, and justify or defy the validity of preassumed spin models. Our work 
not only provides useful insight to understand magnetic electrides
but also suggests a new paradigm to study the related materials.
\end{abstract}


\maketitle

\section{Introduction}

Electrides are a special type of ionic crystals. In these fascinating materials the electrons, trapped in cavity, are the anions. Due to this characteristic feature, many possibly useful properties are realized such as high magnetic susceptibility, low work-function, and strong reducing character, highly variable conductivity, low temperature thermionic emission, high hyperpolarizability \cite{dye_electrides:_2009,yanagi_electron_2012,buchammagari_room_2007,xu_structures_2007}. Recently room-temperature-stable organic and inorganic electride have been synthesized \cite{,satoru_matsuishi_high-density_2003,redko_design_2005}. Most of organic electrides are known to be antiferromagnetic (AFM) from the field response \cite{,redko_design_2005,dawes_cesium_1991,huang_structure_1997,xie_structure_2000,wagner_cs_2000,ichimura_anisotropic_2002,wagner_[cs+15-crown-518-crown-6e-]6_1995}. In the sense that their magnetic properties are presumably originated from the electrons in the cavity, the magnetism of electrides is of unique interest. Although some features of their magnetic interactions have been modeled, such as that the electrons are interacting via a vacant aisle, and classified accordingly \cite{,dye_cavities_1996,dye_electrides:_1997,ryabinkin_solution_2010,ryabinkin_two_2010,ryabinkin_interelectron_2011}, a large part of their fundamental nature still remains elusive. It is largely due to the limitation of conventional ab initio calculation method.

The conventional way of investigating magnetic property from first-principles is to calculate the interaction parameter by comparing multiple total energies corresponding to the ground state and meta-stable magnetic orders. In this way, not only the ground state spin configuration but the magnetic interaction strength are also calculated as shown recently by Dale and Johnson for electrides \cite{dale_explicit_2016}. However, this conventional approach is severely limited when the system size is large which is indeed the case for many organic electrides. 
For large systems, it is difficult to calculate the long-range interactions as the supercell contains too many atoms. While the magnetic interaction in solid is typically classified into the long-range (e.g., Ruderman–Kittel–Kasuya–Yosida (RKKY) interactions) \cite{,ruderman_indirect_1954,yosida_magnetic_1957,kasuya_theory_1956} or short-range (e.g., superexchange interactions) \cite{,kanamori_crystal_1960,goodenough_magnetism_1963} nature, the identification of even such basic character has been hampered by the large unitcell size for organic electrides. On top of their intriguing features of interacting path presumably through some cavity aisle, this practical issue limits the ab initio study. 

In order to meet this challenge, here we introduce a new approach. First we employ so-called ‘magnetic force response theory (MFT)’ \cite{liechtenstein_local_1987,antropov_spin_1996,antropov_exchange_1997,katsnelson_first-principles_2000,han_electronic_2004,yoon_reliability_2018} for calculating magnetic interactions. MFT enables us to calculate all the magnetic interactions residing in a given material within a primitive unitcell and at one time. Thus, without a supercell, one can estimate the magnetic coupling parameter $J$ as a function of distance for both short and long range. In order to understand the magnetic properties of organic electrides and to demonstrate the capability of our computation scheme, we take four different materials; namely, Rb$^{+}$(cryptand[2.2.2])e$^{-}$ (Fig.~\ref{Figure_2}(a)), Li$^{+}$(cryptand[2.1.1])e$^{-}$ (Fig.~\ref{Figure_2}(b)), [Cs$^{+}$(15C5)(18C6)e$^{-}$]$_6$(18C6) (Fig.~\ref{Figure_3}(a)), and  K$^{+}$(cryptand[2.2.2])e$^{-}$ (Fig.~\ref{Figure_4}(a)). Our calculations clearly show that the magnetic interactions in these electrides indeed come from the localized electrons as anions. Further we unveil their short-range versus long-range nature of the interactions. In fact, for some electrides, there is an indication of oscillating $J$ which is a signature of RKKY type magnetic couplings. In order to make MFT feasible, one has to identify the trapped electron states properly. For this purpose, we further employ maximally-localized Wannier function (MLWF) technique \cite{,marzari_maximally_1997,souza_maximally_2001}. Another difficulty in dealing with electrides within first-principles framework is about controlling magnetic order for the cavity electrons. With a special constraint DFT scheme, we successfully stabilized the magnetic solution of K$^{+}$(cryptand[2.2.2])e$^{-}$  for the first time. Our current work provides useful information to understand the magnetism of organic electrides.

\section{ Computational methods }
\subsection{Magnetic force response theory}
MFT is a method to calculate magnetic interactions at a given electronic structure. In this method the exchange coupling is estimated as a response to small spin tiltings as a perturbation from the given converged solution \cite{liechtenstein_local_1987,yoon_reliability_2018}:  
	
	\begin{equation} \label{Eq_Jij_momentumspace}
	J_{ij}({\bf{q}} ) = \frac{1}{\pi} {\rm{Im}} \int_{}^{}  
	\int_{}^{\epsilon_{\rm{F}}}  d{\rm{k}} \, d\epsilon  
	\rm{\, Tr}[
	V_{{\bf{k}},i}^{\downarrow \uparrow } {\mathbf{G}}_{{\bf k},ij}^{\uparrow\uparrow}(\epsilon)   
	V_{{\bf{k+q}},j}^{\uparrow \downarrow}  {\mathbf{G}}_{{\bf k+q},ji}^{\downarrow\downarrow}(\epsilon) 
	].
	\end{equation}
Here i and j are the site indices, and up and down arrow indicate the spin direction. Green's function $\mathbf{G}$ is represented as
	\begin{equation} \label{Eq_green_DFT}
	\mathbf{G}^{\uparrow \uparrow}_{{\bf{k}},ij}(\epsilon) = \sum_{n}^{} \frac{ \ket{\psi_{{\bf{k}},i}^{\uparrow}} \bra{\psi^{\uparrow}_{{\bf{k}},j}} }{\epsilon-\epsilon_{\uparrow n,{\bf{k}}}   + i\eta}
	\end{equation}
where $\epsilon_{n,{\bf{k}}}$ and  $\ket{\psi_{{\bf{k}},i}}$ refers to the n-th eigenvalue and eigenstate, respectively.  $\mathbf{V}$ is given by 
	\begin{equation}\label{Eq_Vdef}
	V_{{\bf{k}},i}^{\downarrow  \uparrow } = {\frac{1}{2}}( {\bf{H}}_{{\bf k},i}^{\downarrow  \downarrow } - {\bf{H}}_{{\bf k},i}^{\uparrow \uparrow } )
	\end{equation}
where ${\bf{H}}_{{\bf k},i}^{\uparrow \uparrow (\downarrow  \downarrow) }$ is Kohn-Sham Hamiltonian corresponding to the collinear up(down) spin.
	
The exchange interaction between two sites is calculated by Fourier transformation from $\bf{k}$ to real space. 
Thus one can just take the minimal size of unitcell with no need to consider large supercells. 
Once the localized magnetic sites are well defined, MFT provides the exchange constants, $J$’s, as a function of distance. 
Note that in MFT we extract all the information from a single self-consistently converged electronic structure. The further details of our implementation and the results of some classical example can be found in our previous studies \cite{,han_electronic_2004,yoon_reliability_2018}. 


\subsection{Calculation details}

We perform density functional theory (DFT) calculations within generalized gradient approximation proposed by Perdew \textit{et al.} (GGA-PBE) \cite{,perdew_generalized_1996} by employing LCPAO (linear combination of pseudo-atomic orbitals) method \cite{,ozaki_variationally_2003,ozaki_numerical_2004} as implemented in our ‘OpenMX’ software package \cite{openmx}. 
It should be noted that the limitation of GGA in describing the correlation effect can cause the overestimation of magnetic couplings or the underestimation of the charge occupation \cite{han_electronic_2004,solovyev_effective_1998,johnson_extreme_2013}. 3×3×3 k-points and 500 Ry energy cutoff are used for numerical integration. 
Poisson equations are solved by using fast Fourier transformations, and the projector expansion method is used to accurately calculate three-center integrals associated with the deep neutral atom potential \cite{ozaki_efficient_2005}.

\begin{figure}[h]
	\begin{center}
		\includegraphics[width=8.6cm,angle=0]{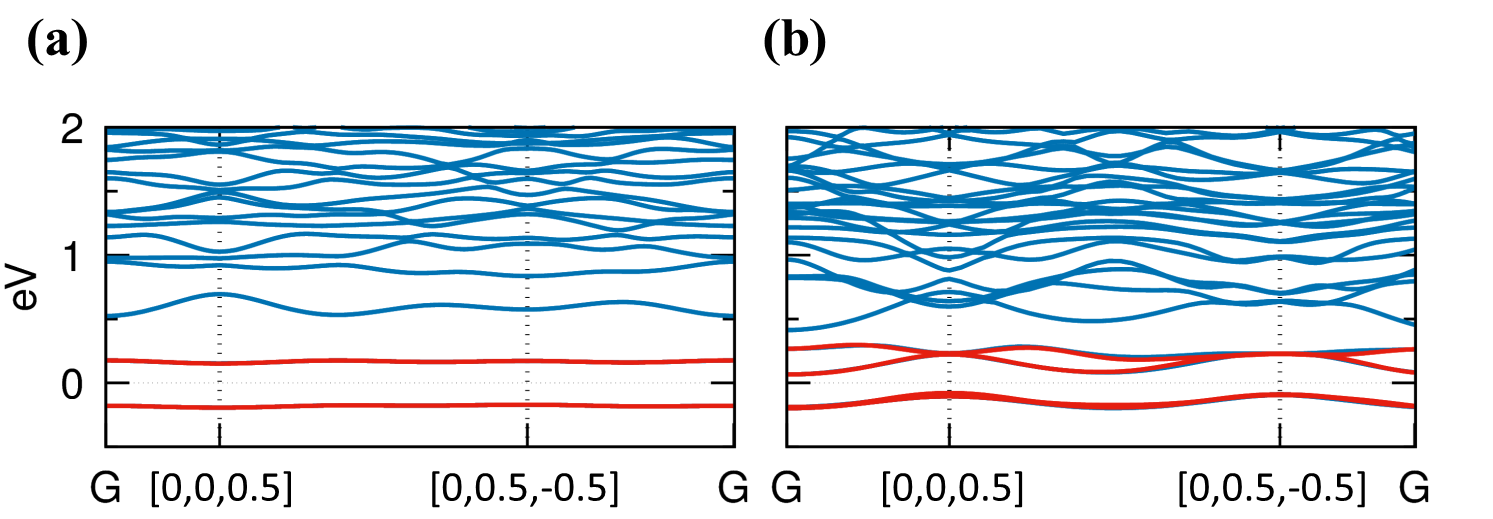}
		\caption{
			(Color online) (a), (b) The calculated band dispersion of (a) Rb$^{+}$(cryptand[2.2.2])e$^{-}$ and (b) Li$^{+}$(cryptand[2.1.1])e$^{-}$ (blue line). The calculated MLWF band dispersion is expressed in red lines. 
			\label{Figure_1}}
	\end{center}
\end{figure}

In order to describe the characteristic feature of electride, namely the electrons trapped in the cavity space, we employ the ‘empty atom’ technique for all of our systems.
The empty atom technique, also called as `ghost atom' or `empty sphere', has been used, for example, to correctly estimate the basis set superposition error (BSSE) \cite{,latajka_dissection_1989,simon_how_1996}, and to treat a large void space \cite{,doi:10.2138/am-1996-11-1201,kalpana_electronic_1996,ossicini_selfconsistent_1989,sque_transfer_2007} within local basis schemes.  In this method, an empty atom is represented as an atom with a nuclear charge of zero, which acts as a basis for describing the wave function of the empty space. We used two $s$, one $p$, and one $d$ orbitals with a cutoff radius of 11 a.u.~ as the basis of empty atom. We determine the number of empty atoms by comparing the band structure with the plane-wave result \cite{,dale_density-functional_2014}. 
MLWF \cite{marzari_maximally_1997,souza_maximally_2001} is also used to identify the magnetic bands as a localized state. We found that the position of Wannier functions was well compared with the distances known from experiments \cite{,huang_structure_1997,xie_structure_2000,wagner_[cs+15-crown-518-crown-6e-]6_1995,dye_electrides:_1990}. This combination of empty atom and MLWF techniques enables us to perform the MFT calculation to estimate the magnetic couplings. 
The s-wave symmetry for Wannier functions is considered to describe the cavity-electron states since they are clearly of s-orbital character from spin density plots seen in Fig.~\ref{Figure_2}(a) and \ref{Figure_2}(b) for example. Figure~\ref{Figure_1}(a) and \ref{Figure_1}(b) shows the calculated band dispersion of Rb$^{+}$(cryptand[2.2.2])e$^{-}$ and Li$^{+}$(cryptand[2.1.1])e$^{-}$, respectively (blue line). The MLWF band is expressed by red lines. An excellent overlap of the two bands (blue and red) shows that MLWF well identifies the electronic states in the cavity.

 MFT calculations are conducted based on the previously-known magnetic phase \cite{,dale_explicit_2016}, namely, G-type AFM order (in which all of the directly connected neighbors have different spins) except for [Cs$^{+}$(15C5)(18C6)e$^{-}$]$_6$(18C6). For [Cs$^{+}$(15C5)(18C6)e$^{-}$]$_6$(18C6), we consider the intra-ring AFM order (see Fig.~\ref{Figure_3}(a)). 
In order to get the AFM phase for all other electrides except K$^{+}$(cryptand[2.2.2])e$^{-}$, the initial  spin polarization is applied to the surrounding hydrogen atoms as in the previous calculation \cite{dale_explicit_2016}.

For K$^{+}$(cryptand[2.2.2])e$^{-}$, however, even such an ad hoc technique does not work as reported in the previous theoretical study \cite{,dale_explicit_2016}. As a result, the magnetic property of this electride has never been theoretically addressed. Here, we used a kind of special constraint DFT technique to stabilize the magnetic order. In this scheme, the initial spin density is assigned onto the ‘empty atom’ sites which means that the constraint is directly imposed on the cavity. We further confirmed that this magnetic solution is not just consistent with experiment but also robust through the further self-consistent steps. It is eventually converged into the magnetic solution without any constraint. From this process, we successfully obtained the well-stabilized self-consistent magnetic solution of K$^{+}$(cryptand[2.2.2])e$^{-}$ for the first time, on top of which MFT can be conducted.

\section{RESULTS AND DISCUSSION}\label{sec:parameter_J}

\subsection{Rb$^{+}$(cryptand[2.2.2])e$^{-}$ and Li$^{+}$(cryptand[2.1.1])e$^{-}$}

Rb$^{+}$(cryptand[2.2.2])e$^{-}$ and Li$^{+}$(cryptand[2.1.1])e$^{-}$ are the electrides with ‘ladder-like’ channel \cite{,huang_structure_1997,xie_structure_2000}. Experimentally, the magnetic property has been studied with electron paramagnetic resonance (EPR) spectroscopy as a function of temperature, and fitting to a certain exchange model.
For these two materials, the EPR data is well fit to ‘FN (first-neighbor)-1D Heisenberg’ model. 
However, a recent calculation study \cite{,dale_explicit_2016} raises a question against this conclusion \cite{comment1}.
The calculated $J$ from the total energy difference mapped onto the same FN-1D model is found to be quite different from the experimental data \cite{huang_structure_1997,xie_structure_2000}. 
It can be argued that this difference is attributed to the next and longer-range magnetic couplings \cite{,dale_explicit_2016}. 
Importantly, however, a solid conclusion could not be made because the next neighbor interactions were not accessible within the conventional computation scheme.

\begin{figure}[t]
	\begin{center}
		\includegraphics[width=8.6cm,angle=0]{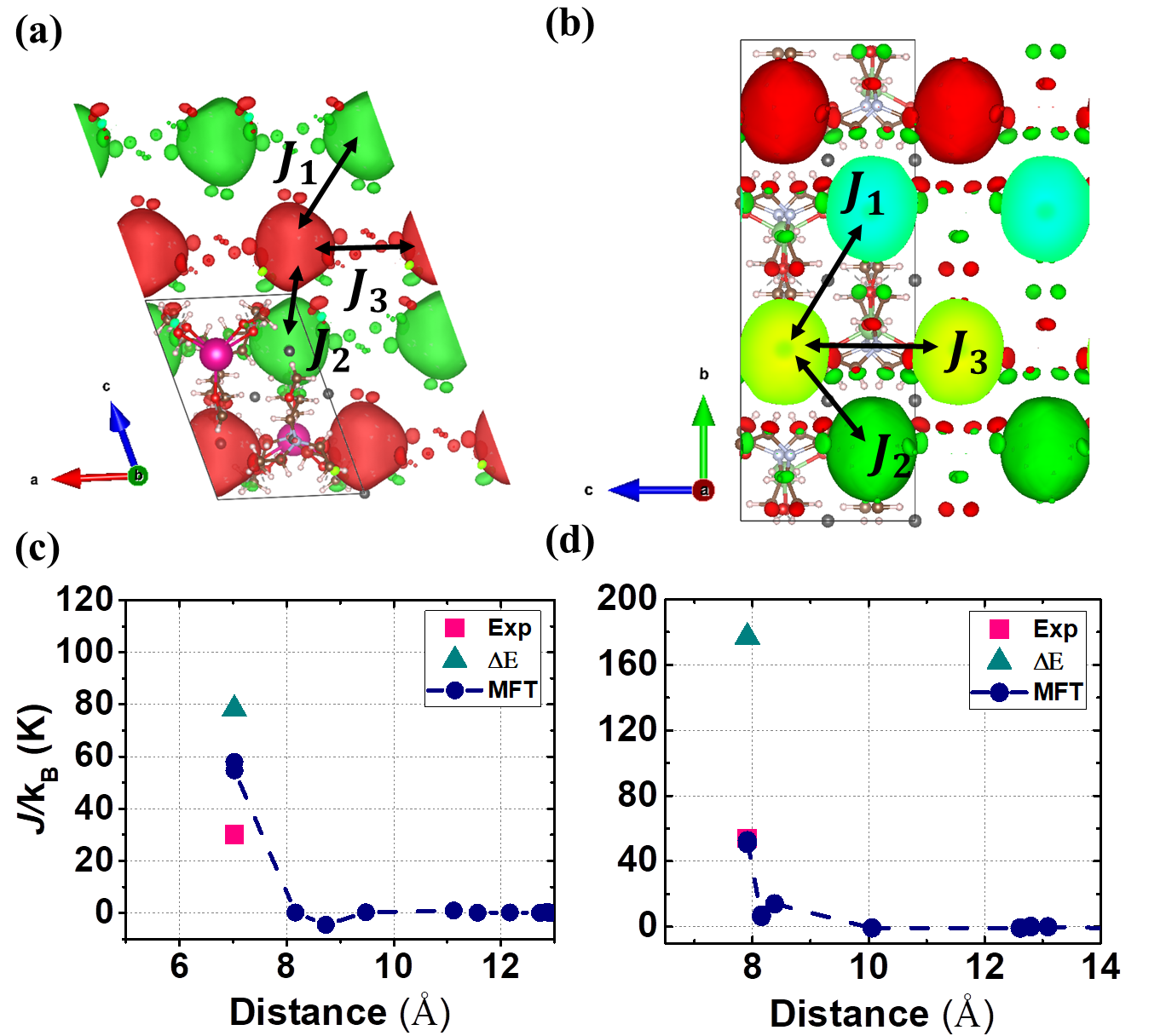}
		\caption{ (Color online) (a), (b) The calculated spin density of (a) Rb$^{+}$(cryptand[2.2.2])e$^{-}$ and (b) Li$^{+}$(cryptand[2.1.1])e$^{-}$ where red and green spheres represent the up and down spin density, respectively. We used the isosurface value of 0.00075 in atomic unit (a.u.).  $J_{1}$, $J_{2}$, and $J_{3}$ refer to the first, second, and third neighbor interactions in (a) and (b). (c), (d) The calculated magnetic coupling parameters for (c) Rb$^{+}$(cryptand[2.2.2])e$^{-}$ and (d) Li$^{+}$(cryptand[2.1.1])e$^{-}$. Our calculation results by MFT (dark blue circles) are compared with the previous calculation by total energy difference (green triangles; Ref. \onlinecite{dale_explicit_2016}) and experiment (magenta squares; Ref. \onlinecite{,huang_structure_1997,xie_structure_2000}). Note that both short and long range interactions are calculated from MFT while only nearest neighbor values can be obtained from experiments and total-energy-based computation scheme.
		}
		\label{Figure_2} 
	\end{center}
\end{figure}

Hereby using MFT combined with MLWF, we calculated the exchange coupling constants as a function of distance; from the nearest neighbor to the long-range interactions. The results are presented in Figs. \ref{Figure_2}(a)-(d). For Rb$^{+}$(cryptand[2.2.2])e$^{-}$ (Fig.~\ref{Figure_2}(a) and \ref{Figure_2}(c)), MFT calculation shows that the first neighbor interaction is dominant, $J_{1}/k_{B}$=56.3 K, while the second and third neighbor interaction are much smaller, $J_{2}/k_{B}$=0.17 K and $J_{3}/k_{B}$= $-$4.56 K (see Table \ref{table1}). Thus our results confirm that Rb$^{+}$(cryptand[2.2.2])e$^{-}$ is well classified as a FN-1D Heisenberg system.

On the other hand, Li$^{+}$(cryptand[2.1.1])e$^{-}$ is not the case (Fig.~\ref{Figure_2}(b) and \ref{Figure_2}(d)). The calculated first neighbor interaction is $J_{1}/k_{B}$=51.7 K, and the second and third neighbor value is quite significant; $J_{2}/k_{B}$=6.69 K and $J_{3}/k_{B}$ =14.0 K (see Table \ref{table1}). Note that this corresponds to 12.9\% and 27.1\% of $J_{1}$, respectively. Thus it is hardly regarded as a FN-1D spin system contrary to the previous study.

Our results demonstrate the usefulness of the current computation scheme for magnetic electrides. The MFT in combination with MLWF provides the long-range interaction parameters and therefore one can have a reliable picture for the magnetism even if the system size is too large to be calculated by conventional total energy method. Further, our calculation confirms that the magnetic properties measured in the experiment indeed come from the localized electron state in the cavity sites. It is because MFT estimates the magnetic force in between two sites, and in the current case, these ‘sites’ are defined as the cavity electrons by means of empty atoms and MLWF technique. We also emphasize that, in our scheme, one does not need to build any a priori model considering only a few neighbor interactions. The exchange parameters are calculated as the response to spin tilting. This feature is advantageous when one needs to consider any type of model building or to justify a certain model.

\subsection{[Cs$^{+}$(15C5)(18C6)e$^{-}$]$_6$(18C6)}

\begin{figure}[!ht]
	\begin{center}
		\includegraphics[width=8.6cm,angle=0]{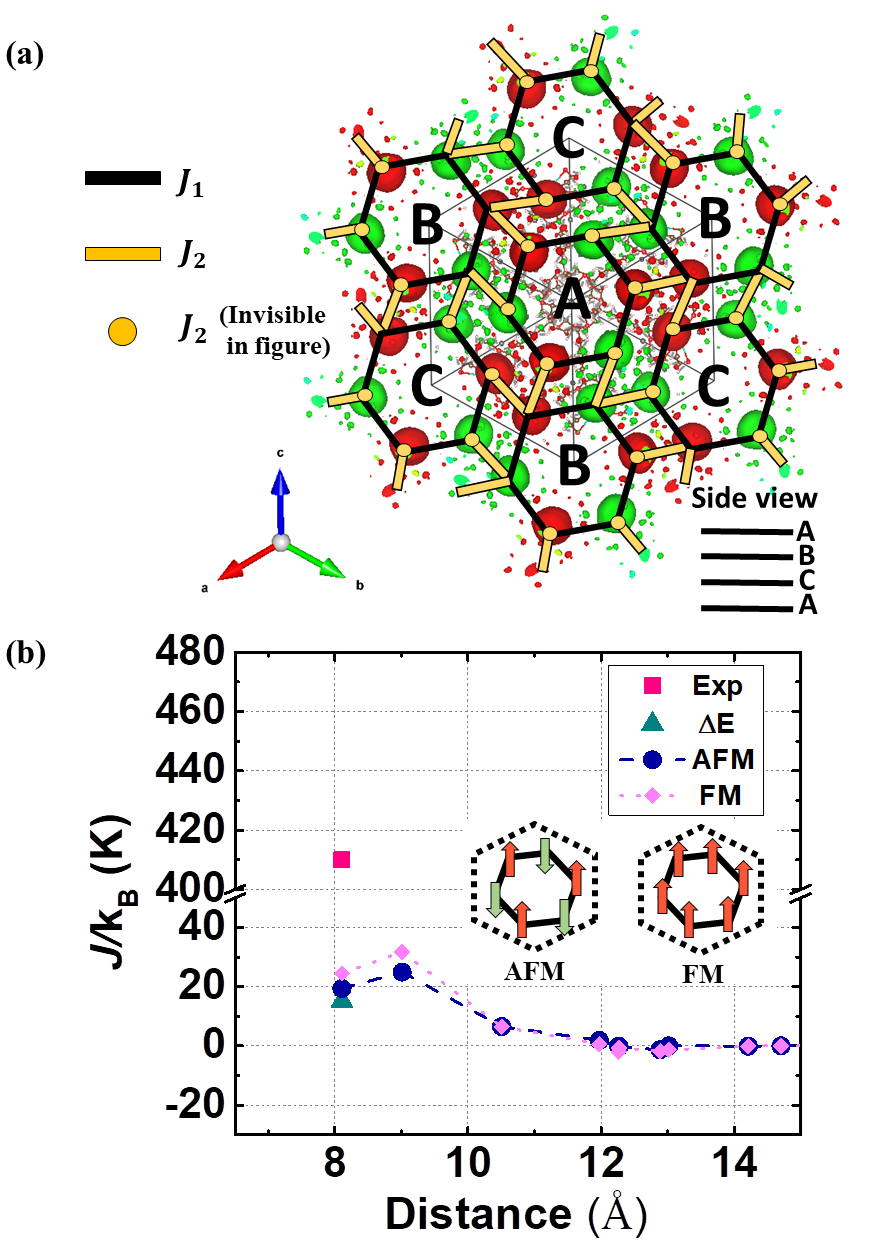}
		\caption{ (Color online) (a) The calculated spin density of [Cs$^{+}$(15C5)(18C6)e$^{-}$]$_6$(18C6) where red and green spheres represent the up and down spin density, respectively. We used the isosurface value of 0.00075 a.u.. The spin order in this material can be referred as ‘intra-ring antiferromagnetic (intra-ring AFM)’ configuration: Black lines show the connections of the intra-vacant spaces of ‘six-memebered rings’ while yellow lines show the connections between the different ‘six-memebered rings’. Yellow points represent the connections which are invisible in the current view of Figure. Note that each site is connected by two black lines and two yellow lines or points. In terms of distance, the black lines correspond to the first neighbor interactions ($J_{1}$) and the yellow dots/lines to the second neighbors ($J_{2}$). The stacking sequence of the six-membered rings is A-B-C-A-… from the side view. 
			(b) The calculated exchange interaction for [Cs$^{+}$(15C5)(18C6)e$^{-}$]$_6$(18C6). We estimated $J$ values based on two different spin density configurations (see main text for more details), and the results are presented with dark-blue circles and pink diamonds. For comparison, the previous calculation (Ref. \onlinecite{,dale_explicit_2016}) and experimental (Ref. \onlinecite{,wagner_[cs+15-crown-518-crown-6e-]6_1995}) data are also presented. The two inset figures show the schematic spin configuration based on which we performed the MFT calculation for $J$; namely, intra-ring AFM and ferromagnetic (FM) order.
		}
		\label{Figure_3} 
	\end{center}
\end{figure}

[Cs$^{+}$(15C5)(18C6)e$^{-}$]$_6$(18C6) is an electride having ‘six-membered ring’ cluster structure as shown in Fig.~\ref{Figure_3}(a), and its magnetic susceptibility was reported by Wagner and Dye \cite{,wagner_[cs+15-crown-518-crown-6e-]6_1995}. We note that, in this experimental study, the measured data has been analyzed based on a simple model assumption that the intra-ring couplings are the only magnetic interactions \cite{,wagner_[cs+15-crown-518-crown-6e-]6_1995}. From this analysis, it was concluded that this electride has strong AFM interactions.

The theoretical study was also conducted under the same assumption that only the first-neighbor interaction is important \cite{,dale_explicit_2016}. It is presumably because the computational cost is too large to calculate multiple total energies within the enlarged supercell geometry. In fact, in order to simulate the theoretically-estimated G-type spin ground state, one needs to calculate the 4080-atom supercell. In Ref.~\onlinecite{,dale_explicit_2016}, the cell size of 510 atoms has been used within which only the nearest neighboring $J$ is accessible. This result is plotted in Fig.~\ref{Figure_3}(b) (denoted by green triangle) showing that the difference between the previous calculation and experiment is about 395 K. 
The origin of the relatively large difference is unclear.
It was indeed concluded in Ref.~\onlinecite{,dale_explicit_2016} that this difference is attributed to the supercell size effect.

In this context, MFT can give a useful insight since it does not require any supercell calculation but it does provide the longer range interactions. Fig.~\ref{Figure_3}(b) shows our MFT calculation results of exchange couplings as a function of distance (see dark-blue circles and pink diamonds). Since the magnetic ground state cannot be represented within the structural primitive cell, we considered two different magnetic solutions which can be realized within this primitive cell; namely, FM (pink diamonds) and AFM (dark-blue circles). Here AFM order refers to the AFM order in between the intra-rings (see the inset of Fig.~\ref{Figure_3}(b)). 
We first note that the two calculation results based on FM and AFM spin density are quite similar; the difference is less than 6.5 K. 
It not only indicates that the spin-polarized electron states in cavity are basically well localized as previously discussed \cite{,yoon_reliability_2018}, but also implies that we do not need the real spin ground state density in order to estimate $J$ values \cite{,yoon_reliability_2018}.

\begin{table*}[t]	
	\centering
		\begin{tabular}{|c|c|c|c|c|c|c|c|c|c|c|c|c|c}
			\hline
			\hline
			\backslashbox{ $J_{n}/k_{B} $}{ Name} & \multicolumn{3}{c|}{ Rb$^{+}$(cryptand[2.2.2])e$^{-}$} & \multicolumn{3}{c|}{ Li$^{+}$(cryptand[2.1.1])e$^{-}$} & \multicolumn{3}{c|}{
				 [Cs$^{+}$(15C5)(18C6)e$^{-}$]$_6$(18C6)} & \multicolumn{3}{c|}{ K$^{+}$(cryptand[2.2.2])e$^{-}$} \\
			\hline
			&  \enspace MFT \enspace & \enspace $\Delta$E\enspace &  Exp    
			&  \enspace MFT \enspace & \enspace $\Delta$E\enspace &  Exp        
			&  \enspace MFT \enspace & \enspace $\Delta$E\enspace &  Exp        
			&  \enspace MFT \enspace & \enspace $\Delta$E\enspace &  Exp  \\
			\hline
			$J_1$ & 56.3 &  78.2  & 30  & 51.7 & 177 & 54 & 19.4&15.2  & 410 & 9.6 & - &  440 \\ 
			\hline
			$J_2$& 0.2 & - & - & 6.7  & - & - & 24.9  & -& -& 5.7 &  - &  - \\  
			\hline
			$J_3$& $-$4.6  & - & -  & 14.0  & -& - & 6.6  & - & - & 148 & - & - \\ 
			\hline
			$J_7$& 0.1  &  -  & -  & 0.02  & -  & -  & $-$1.3 & - & - & $-$33.5  & -  & - \\ 
			\hline
		\end{tabular}
	\caption{\label{table1} 
	Exchange parameters of 4 different organic electrides in the unit of K. 
	Our calculation results by MFT are compared with the previous calculations ($\Delta$E; Ref.~\cite{dale_explicit_2016})
	and experiments (Exp). The experimental data can be found in 
	Ref.~\cite{xie_structure_2000},
	Ref.~\cite{huang_structure_1997}, 
	Ref.~\cite{wagner_[cs+15-crown-518-crown-6e-]6_1995} and 
	Ref.~\cite{,ichimura_anisotropic_2002} 
	for Rb$^{+}$(cryptand[2.2.2])e$^{-}$,
	Li$^{+}$(cryptand[2.1.1])e$^{-}$,
	[Cs$^{+}$(15C5)(18C6)e$^{-}$]$_6$(18C6) and
	K$^{+}$(cryptand[2.2.2])e$^{-}$, respectively.
	}

\end{table*}

The sum of all magnetic interactions is $J_{tot}/k_{B}$ = ($J_{1}$+$J_{2}$+… )$/k_{B}$= 53 K which is notably smaller than the experimental value of 410 K (magenta square) \cite{wagner_[cs+15-crown-518-crown-6e-]6_1995} (see Table \ref{table1}). It means that the difference between the previous calculation and experiment is not originated from the longer-range interactions or supercell-size effects. 
Comparing our MFT result with the previous total energy-based estimation, the first neighbor $J_{1}$ is in good agreement with each other within 4.2 K. Importantly, the second neighbor $J_{2}$ is comparable with and slightly larger than $J_{1}$. Namely, the inter-ring interaction is sizable and the nearest neighbor spin model is not relevant to this material. 
Our calculations  clearly show that the inter-ring coupling needs to be taken into account.

Further, our calculation confirms that the spin ground state of this material is indeed G-type.
As mentioned above, this material was speculated to have G-type AFM ground state \cite{dale_explicit_2016}. However, there is no experimental or theoretical evidence for that.
Theoretical confirmation has been hampered by the large supercell size of 4080 atoms.
Our MFT results of Fig.~\ref{Figure_3}(b) clearly shows that $J_1$ and $J_2$ are the two dominant couplings and both of them are AFM.

\subsection{K$^{+}$(cryptand[2.2.2])e$^{-}$}

\begin{figure}[!t]
	\begin{center}
		\includegraphics[width=8.5cm,angle=0]{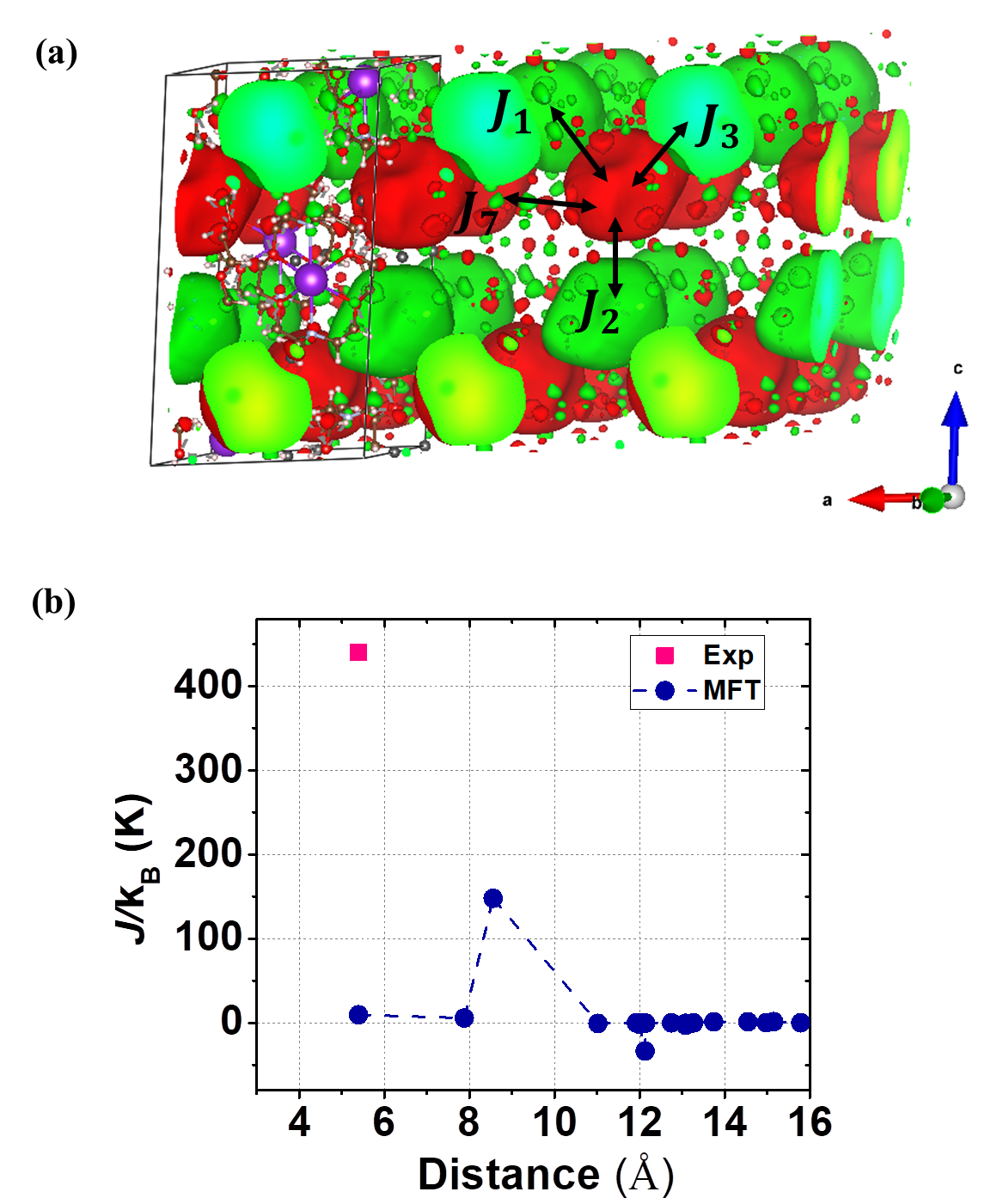}
		\caption{ 
			(Color online) (a) The calculated spin density of K$^{+}$(cryptand[2.2.2])e$^{-}$ where the red and green spheres represent the up and down spins, respectively. We used the isosurface value of 0.0003 a.u.. (b) The calculation results of magnetic interaction as a function of distance. The magenta square shows the experimental result (Ref. \onlinecite{ichimura_anisotropic_2002}) while the dark-blue circles are our calculation results. For this material, there is no previous calculation result because of the difficulty in stabilizing the magnetic solution (see the main text for more details).
		}
		\label{Figure_4}
	\end{center}
\end{figure}

Our final example, K$^{+}$(cryptand[2.2.2])e$^{-}$, is also known to be an AFM ordered organic electride \cite{,ichimura_anisotropic_2002}. For this material, however, there has been no successful calculation in obtaining the magnetic solution. The conventional DFT calculation gives the paramagnetic spin ground state as the converged solution even if starting from the spin polarized initial condition. Even with ad hoc treatment in which initial spins are assigned to the hydrogen atoms around the vacant space, the magnetic solution is hardly achieved \cite{,dale_explicit_2016}.

First of all, we successfully obtained the magnetic solution by applying initial spins directly on the vacant space. 
We developed a constrained	DFT scheme for assigning the initial spin moment to `empty	sphere'. 
This process is not straightforward in the sense that electron spins need to be polarized within the empty spheres.
We applied the magnetic constraint during the initial $~20$ self-consistent steps, through which empty spheres are occupied by polarized electrons. 
After that, the usual DFT self-consistent calculations are performed with the constraint turned off. 
With this scheme, we successfully generated the well-stabilized AFM solution as shown in Fig.~\ref{Figure_4}(a).

On top of this spin density we performed MFT calculation and successfully estimated magnetic interactions for the first time. Our results are summarized in Fig.~\ref{Figure_4}(b). First of all, our calculation confirms that this magnetic phase is
indeed G-type AFM ground state as speculated in the previous study \cite{dale_explicit_2016}.
It is also consistent with an experiment \cite{ichimura_anisotropic_2002}.
The largest interaction is the third neighbor $J_{3}$ which is notably larger than $J_{1}$, $J_{2}$ and others.
Interestingly  the second largest interaction is $J_{7}$ which is about 25\% of the $J_{3}$.
Our results is consistent with the experiment \cite{ichimura_anisotropic_2002} in the sense that the magnetic property of this material can be described with two parameters in two dimension.
The total sum of all our $J/k_{B}$’s is about 110 K (see Table \ref{table1}). 


\section{Summary}\label{sec:MFT_metastable}

A new theoretical approach is applied to study magnetic electrides. Spin-polarized electrons trapped in the cavity are identified by empty atom and MLWF method, and their interactions calculated within MFT. 
The usefulness of this scheme is  shown by calculating four different organic electrides for which the validity of pre-assumed models have remained  unclear. 
The long range magnetic interaction profile as a function of distance is calculated and compared, which has not been accessed by the conventional total energy calculations. 
 For K$^{+}$(cryptand[2.2.2])e$^{-}$, we apply a constraint DFT method to stabilize the magnetic solution and  calculate the magnetic interaction for the first time. Our study provides useful insights to understand magnetic electrides and related materials.

\section{Acknowledgments}
This work was supported by Basic Science Research Program through the National Research Foundation of Korea (NRF) funded by the Ministry of Education (2018R1A2B2005204).

\bibliography{Organic_electride_v8}

\end{document}